# Hong-Ou-Mandel dip using photon pairs from a PPLN waveguide

Qiang Zhang[1], Hiroki Takesue[2], Carsten Langrock[1], Xiuping Xie[1], M. M. Fejer[1], Yoshihisa Yamamoto[1,3]

*1. Edward L. Ginzton Laboratory, Stanford University, Stanford, California 94305*
*2. NTT Basic Research Laboratories, NTT Corporation, 3-1 Morinosato Wakamiya, Atsugi, Kanagawa 243-0198, Japan*
*3. National Institute of Informatics, 2-1-2 Hitotsubashi, Chiyoda-ku, Tokyo, 101-843, Japan*
qiangzh@stanford.edu

**Abstract:** We experimentally observed a Hong-Ou-Mandle dip with photon pairs generated in a periodically poled reverse-proton-exchange lithium niobate waveguide with an integrated mode demultiplexer at a wavelength of 1.5 μm. The visibility of the dip in the experiment was 80% without subtraction of any noise terms at a peak pump power of 4.4 mW. The new technology developed in the experiment can find various applications in the research field of linear optics quantum computation in fiber or quantum optical coherence tomography with near infrared photon pairs.

## 1. Introduction

When two identical photons overlap at a beam splitter from different inputs, the two photons will bunch together upon either beam splitter output and no coincidence of the two outputs will be observed. This is the so-called Hong-Ou-Mandle (HOM) dip [1]. It was first used for precisely measuring the time duration of wide-spectral-bandwidth single photons [1], which were too faint to be measured using an autocorrelator. Classical light fields also exhibit a similar destructive interference, but only with a 50% visibility. Therefore, the HOM dip is often utilized to demonstrate the difference between the regime of quantum optics and its classical counterpart [2]. Very recently, the effect found several important applications in the field of quantum information, for example, quantum teleportation [3], quantum repeater [4], linear optics quantum computation (LOQC) [5], and quantum optical coherence tomography (QOCT) [6], among others.

The HOM dip was first observed using visible-light photon pairs generated in a nonlinear crystal via parametric down-conversion (PDC) [1]. Soon afterwards, it was also realized using telecom-band photon pairs, either generated via PDC [7], or via four-wave mixing in dispersion-shifted [8] or photonic crystal fibers [9].

As mentioned earlier, conventional PDC is most often implemented using bulk nonlinear crystals. Recently, periodically poled lithium niobate (PPLN) waveguides have been used to generate correlated photon pairs due to their higher conversion efficiency compared to their bulk counterparts [10]. Furthermore, waveguide devices can be fiber pigtailed to achieve high coupling efficiencies, as well as simple and mechanically stable operation. However, the HOM dip has not yet been observed using photon pairs from PPLN waveguides due to the difficulty of spatially separating identical photon pairs. In the fiber-optic implementation, somewhat complicated methods to separate the identical photon pairs are considered [11].

In this letter, we used an asymmetric Y-junction-based mode demultiplexer realized in a reverse-proton-exchange (RPE) PPLN waveguide [12,13] to separate the degenerate photon pairs generated in the quasi phasematching (QPM) grating. The photon pairs were then launched into a fiber-based beam splitter and a 80%-visibility HOM dip was observed with an average of 0.03 photon pairs per pulse without subtraction of any accidental coincidences.

## 2. System

PDC processes in nonlinear crystals usually utilize angular or polarization-state differences between the signal and the idler photon to separate them. However, our protonated z-cut $LiNbO_3$ waveguides support only TM-polarized waves, eliminating the possibility of polarization de-multiplexing. Naturally, angular de-multiplexing is impossible in a channel waveguide configuration.

Mode de-multiplexing using asymmetric Y-junctions, based on the adiabatic variation of the refractive index distribution along the device, can be utilized to solve this problem. Both the input and output sides have two arms, one narrow and one wide, as shown in Figure 1. The pump wave in the $TM_{00}$ mode is coupled into the narrow input arm of the Y-junction and is adiabatically converted into the $TM_{10}$ mode of the main waveguide. The pump then traverses the QPM grating and generates photon pairs in different spatial modes (signal in $TM_{10}$ and idler in $TM_{00}$, respectively). The idler photon in the $TM_{00}$ mode will exit via the wide arm of the Y-junction's output port, while the signal photon in the $TM_{10}$ mode will be coupled into the narrow arm. Both output modes are then adiabaticaly transformed to $TM_{00}$ modes with identical sizes. For wavelengths near 1.5 μm, an extinction ratio $\xi$ >30 dB between the two output modes can be achieved with current fabrication technologies [12,13]. In the past, we have utilized this type of waveguide to generate 10-GHz-clock-rate photon pairs [14].

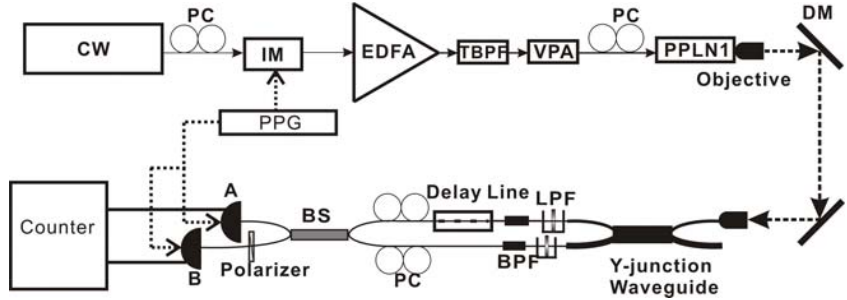

Fig. 1. Diagram of the experimental setup; IM: intensity modulator, VPA: variable power attenuator, A and B: InGaAs APDs, BS: beam splitter.

In the experiment, an external-cavity diode laser operating at a wavelength of 1559 nm was modulated by an optical intensity modulator into pulse trains with 100-MHz repetition rate and 90-ps pulse width. As shown in Figure 1, the 100-MHz signal for the amplitude modulator was derived from a pulse-pattern-generator (PPG). After being amplified by an erbium-doped fiber amplifier (EDFA), the laser pulses with an average power of 11.48 mW passed through a 3-nm-wide tunable bandpass filter (TBPF) to reduce the spontaneous emission noise from the EDFA. A variable attenuator was inserted after the bandpass filter to regulate the pump power. The pump was then frequency doubled in the first PPLN waveguide chip. Since these waveguide devices only accept TM-polarized light, an in-line polarization controller was used to adjust the polarization. The residual pump was attenuated by 180 dB using dichroic mirrors (DM). The second harmonic (SH) wave was then launched into the second asymmetric-Y-junction-containing PPLN waveguide as the pump for the PDC process. The two output ports were fiber pigtailed to improve the collection efficiency and stability of operation. A fiber U-bench (Thorlabs FB220-FC) with several long-pass filters (LPF) was inserted into each output fiber of the second PPLN chip to eliminate residual SH and other nonlinear fluorescence components. The second PPLN waveguide was operated at degeneracy by adjusting the crystal temperature. The bandwidth of the down-converted photon pairs was 40 nm, such that the group velocity dispersion in the PPLN waveguide could not be ignored. Two 0.8-nm-wide BPFs were used to reduce the spectral width and define the photon pairs' coherence time as 4 ps (FWHM).

The photon pairs were combined at a 50:50 fiber beam splitter. To achieve perfect temporal overlap, the signal photons passed through a variable fiber-coupled free-space delay line before being launched into the beam splitter. The minimum step size of the delay line was about 30 fs, which was much smaller than the photon pairs' coherence time, 4 ps. We inserted a polarization controller into each channel and one polarizer before detector B to equalize the polarization of the two photons.

We used two InGaAs avalanche photodiodes (APDs) operated in the gated mode with a 5-MHz gate frequency to detect the output photons of the beam splitter. The gate pulses were synchronized to the incoming photons via the PPG. The quantum efficiency and dark count rates of the detectors were 10% (both), 544 Hz and 1596 Hz, respectively. The two detectors were connected to a counter for coincidence measurements.

## 3. Theory and Result

When the pump-pulse duration is longer than the photon-pair coherence time in the PDC process, which was the case in our experiment, the pair generation probability obeys Poissionian statistics [15,16]. The photon pair's quantum state can then be expressed as:

$$|\psi\rangle = (1 - p - p^2)|0\rangle + p|1\rangle_s|1\rangle_i + \frac{p^2}{2}|2\rangle_s|2\rangle_i + O(p^2), \quad (1)$$

where $p$ is the average photon-pair number per pulse, $s, i$ represent the signal and idler channel, $|1\rangle$ or $|2\rangle$ represent the photon-number Fock state.

The parameter "$p$" is very important for analyzing the visibility of the HOM dip and can be estimated by looking at the coincidence accidental ratio (CAR) [14]. To measure the CAR, we slightly modified the experimental setup in Fig. 1 as follows. We exchanged the counter for a time interval analyzer (TIA) and purposely misaligned the fiber delay's position away from the HOM dip condition. The detection events of the signal and idler channels were used as start and stop signals of the TIA, respectively [14]. A coincidence in the matched time slot represents the true coincidence caused by photons generated with the same pump pulse, while a coincidence in the unmatched time slot represents the accidental coincidence mainly caused by photons generated by different pump pulses. The ratio between the two coincidence rates is the CAR. In our experiment, the CAR reached 29.4 at a peak pump power of 4.4 mW, corresponding to an average photon-pair number of ~0.03 per pulse.

Since the spectral shape of the 0.8-nm-wide BPFs was Gaussian, the shape of the photon pairs was also approximated by a Gaussian. In this case, the coincidence rates $N$ of the HOM dip can be fitted by the following function [7,8],

$$N = C\left\{1 - \frac{2VRT}{R^2 + T^2}\exp(-\frac{\delta\tau^2}{2\sigma^2})\right\} \quad (2)$$

where $C, V, T, R, \delta\tau, \sigma$ represent the coincidence rates away from the HOM dip, visibility, the transmittance and the reflectance of the 50:50 coupler, the delay time, and the $1/e$ temporal half-width of the photon field. In our case, $T$ and $R$ were –3.3 and –3.6 dB, respectively. $\sigma$ was 1.7 ps according to the 4-ps FWHM of the photon pair duration. As mentioned earlier, the average photon-pair number per pulse "$p$" was ~0.03, leading us to neglect higher-order terms ($>p^2$) in the evaluation of the visibility $V$. In the experiment, several effects contributed to the accidental coincidences in the HOM dip. For multi-photon-pair emission, we only considered terms up to $p^2$, while for the dark count contribution we only considered one detector being triggered accidentally and the other by an actual photon. The visibility can then be expressed by:

$$V = 1 - \frac{2p\eta + 4Dt + \eta/\xi}{\eta + 3p\eta + 4Dt + \eta/\xi}, \quad (3)$$

where $\eta, D, t, \xi$ are the channel loss (including the detector's quantum efficiency), the dark count rate, the coincidence time window, and the extinction ratio, respectively.

We set the average pulse peak pump power to 4.4 mW by adjusting the variable power attenuator and scanned the fiber delay to observe the HOM dip shown in Figure 2(a). From a fit to the data, we estimated the visibility to be $(80 \pm 9)$% and $\sigma$ as $1.20 \pm 0.20$ ps, close to the theoretical estimate of 1.7 ps. Thus, a visibility that indicates quantum interference (>50%) was observed [2].

In order to show the relation between the visibility and the average photon-pair number per pulse, we varied the pump power and measured the corresponding visibility of the HOM dips. From Figure 2(b), we can see that the visibility is inversely proportional to the average photon-pair number.

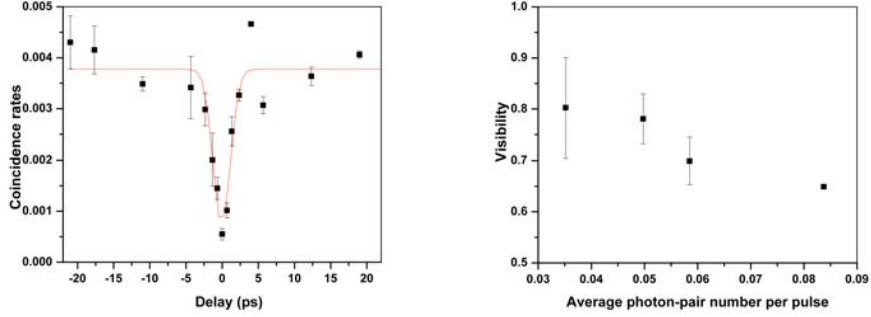

Fig. 2(a). HOM dip. We took three sets of data for each point in the graph to estimate the measurement uncertainty. The error bars result from the data sets' standard deviation. (b). HOM dip visibility with versus average photon-pair number per pulse.

## 4. Conclusion

We generated frequency-degenerate but spatially separate twin photons and observed HOM dip using a simple setup, a RPE PPLN waveguide using asymmetric Y-junction-based mode multi-/demulti-plexer. The HOM dips observed in the experiment originate from quantum interference and are consistent with our theoretical analysis. We are confident that this integrated photon-pair source can find applications in QOCT and LOQC.

**Acknowledgement**

This research was supported by NICT, the MURI center for photonic quantum information systems (ARO/ARDA program DAAD19-03-1-0199), SORST, CREST programs, Science and Technology Agency of Japan (JST), the U.S. Air Force Office of Scientific Research through contracts F49620-02-1-0240, the Disruptive Technology Office (DTO). We acknowledge the support of Crystal Technology, Inc.